# Uncovering Phase Change Memory Energy Limits by Sub-Nanosecond Probing of Power Dissipation Dynamics

Keren Stern[1,†], Nicolás Wainstein[1,†], Yair Keller[1], Christopher M. Neumann[2], Eric Pop[2,3], Shahar Kvatinsky[1], and Eilam Yalon[1,*]

[1]*Electrical Engineering, Technion - Israel Institute of Technology, Haifa, 32000, Israel*

[2]*Electrical Engineering, Stanford University, Stanford, CA 94305, USA.*

[3]*Materials Science and Engineering, Stanford University, Stanford, CA 94305, USA.*

[†]These authors contributed equally

[*]Contact: eilamy@technion.ac.il

**Phase change memory (PCM) is one of the leading candidates for neuromorphic hardware and has recently matured as a storage class memory. Yet, energy and power consumption remain key challenges for this technology because part of the PCM device must be self-heated to its melting temperature during reset. Here, we show that this reset energy can be reduced by nearly two orders of magnitude by minimizing the pulse width. We utilize a high-speed measurement setup to probe the energy consumption in PCM cells with varying pulse width (0.3 to 40 nanoseconds) and uncover the power dissipation dynamics. A key finding is that the switching power ($P$) remains unchanged for pulses wider than a short thermal time constant of the PCM ($\tau_{th}$ < 1 ns in 50 nm diameter device), resulting in a decrease of energy ($E=P\cdot\tau$) as the pulse width $\tau$ is reduced in that range. In other words, thermal confinement during short pulses is achieved by limiting the heat diffusion time. Our improved programming scheme reduces reset energy density below 0.1 nJ/µm$^2$, over an order of magnitude lower than state-of-the-art PCM, potentially changing the roadmap of future data storage technology and paving the way towards energy-efficient neuromorphic hardware.**

## 1. INTRODUCTION

Phase-change memory (PCM) stands as a promising candidate for storage,[1,2] neuromorphic computing,[3] radio-frequency applications,[4,5] and in-memory computing,[6–8] thanks to its nonvolatility, long retention, high endurance, short switching time, and compatibility with the back-end of line (BEOL) of standard silicon processing.[9] Intel's Optane memory, a PCM-based memory aimed at storage class memory applications,[10] is an example of the maturity of this technology. PCM is a two-terminal nonvolatile memory that relies on the high resistivity ratio ($>10^3$) between the amorphous and crystalline phase of chalcogenides glasses, mainly compounds of the ternary $Ge_xSb_yTe_z$. The phase transition is triggered by Ovonic threshold switching (OTS),[1,11] where the resistance of the chalcogenide is reduced above a certain threshold voltage, enabling high current densities and self-heating for crystallization.

The memory cell structure consists of a phase change material, such as $Ge_2Sb_2Te_5$ (GST), sandwiched between two electrodes. Electrical current crowding is achieved by either a small-area heater bottom electrode (BE) in a mushroom-type cell, or a narrow pore in the surrounding dielectric of the PCM layer in a so-called confined cell.[2] To crystallize the PCM (*i.e.*, set) to its low resistance state (LRS), a medium-amplitude voltage or current pulse is applied, which heats the material above its crystallization temperature. The set pulse must be sufficiently long to allow for the material to crystallize. To amorphize the cell (*i.e.*, reset) to its high resistance state (HRS), a large-amplitude electrical pulse is applied for a short time (or with a short fall time) to melt the material and rapidly quench it below the crystallization temperature.[2,12,13]

Because the phase transition is thermally induced, device-level heat management is crucial. The energy and power consumption are of great concern, and their fundamental limits are yet to be fully understood.[14] Particularly, the reset process includes heating the phase change material above its melting temperature (typically $T_m > 600$ºC) which requires very high power density, on the order of ~10 MW/cm$^2$ (100 mW/µm$^2$). Reports on the values of reset power (during the pulse) and energy are scarce because of the need to capture the current and voltage during a short transient (typically less than 100 ns), but the reported energies per area are in the range of ~ 1 J/cm$^2$ (10 nJ/μm$^2$).[15,16] Previous research efforts towards improving energy-efficiency (heating energy per device area) of the melt-quench process in PCM focused on thermal engineering of materials,[17,18] interfaces,[19,20] and device structure.[21,22] However, the range of thermal properties of materials and their interfaces is limited.[23]

In this article, we probe the power dissipation dynamics of PCM reset with sub-nanosecond (ns) resolution and show that the energy efficiency improves with the reduction of programming pulse width, *e.g.*, the energy consumption can be reduced by approximately 100× as the pulse-width (PW) is reduced from 40 ns to 0.4 ns. Although sub-ns reset in PCM was already shown in several studies,[24–28] the power and energy consumption during the sub-ns pulse application were not characterized to date. Measuring the current and voltage across the PCM during a sub-ns pulse is challenging because of parasitic capacitance and signal reflections at high frequencies (>1 GHz) in conventional on-wafer probing.

To circumvent this challenge, here we utilize a high-speed measurement setup, enabling us to directly probe the power and energy consumption in sub-ns reset pulses by measuring the current and voltage across the PCM during the applied pulse. We show that the reset power remains unchanged while reducing the PW below 5 ns for devices with via size in the range 50 to 200 nm. The reset power starts to increase only for PW shorter than the PCM thermal time constant ($\tau_{th}$), namely ~2-3 ns and sub-1 ns for 200 nm and 50 nm nominal via size, respectively. As a result, the optimal energy-power-efficiency point can be traced. Although the power consumption increases as the PW is reduced below $\tau_{th}$, the energy consumption continues decreasing with reduced PW, within the measurement limit of our experimental setup (0.3 ns). We also perform finite-element method (FEM) simulations which predict the minimum reset power and energy, in good agreement with our experimental results. In addition to the characterization of power and energy consumption, our high-speed setup also allows us to study the thermal dynamics of the PCM. A drop in the transient resistance of the device during reset can be associated with melting of a critical volume of the GST, *e.g.*, $T > T_m$ (where $T_m$ is the temperature required to melt the GST). Measurements of the transient resistance therefore allow us to determine the thermal time constant of the device, which is correlated with the optimal pulse for power-efficiency. Finally, reset pulses with the shortest PW of 0.3 ns (limited by instrumentation) achieve record-low energy consumption per area, namely ~0.2 pJ for confined PCM cells with ~50 nm via diameter or ~0.1 nJ/μm$^2$, nearly two orders of magnitude below typical PCM reset energy density.[15]

## 2. RESULTS AND DISCUSSION

### 2.1. High-speed measurement setup and device structure

The measurement setup (**Figure 1a**) consists of a fast pulse generator (PG) connected in series with the PCM device and a high-speed oscilloscope (see Methods). To support sub-nanosecond pulses, ground-signal (GS) transmission lines with characteristic impedance $Z_0 = 50\ \Omega$ were used to connect the device with the measurement equipment. Transmission lines reduce pulse broadening, signal reflections, and possible ringing due to parasitic capacitances. The output impedance of the PG, as well as the input impedance of the scope are also set to 50 Ω.

In this work, we focus on confined PCM cells as shown in **Figure 1b,c**, but the results are not limited to a specific PCM structure. The devices were fabricated as follows. First, tungsten (W) was evaporated, patterned, and etched to form the bottom electrode (BE). Next, $SiO_x$ was deposited using plasma enhanced chemical vapor deposition (PECVD) and the confined vias were patterned using e-beam lithography. Sputtering and lift-off were used to pattern the $Ge_2Sb_2Te_5$ (GST) layer with TiN capping as well as the final TiN/Pt top electrode (TE) and contact pads. More fabrication details are given in the Methods Section. The confined vias have different diameters, nominally: 50 nm, 75 nm, 100 nm, and 200 nm, and the contact area of each set of devices was imaged using scanning electron microscopy (SEM) in order to measure their actual size, more details can be found in Figure S1 and Table S1.

The devices are initially in the LRS (*i.e*. set, crystalline) of few kΩ (**Figure 1d**). We measure the PCM voltage and current during pulse application (**Figure 1e**) to obtain the transient resistance, power, and energy consumption. Following a successful reset pulse, the device is switched to the HRS (*i.e*. amorphous) of hundreds of kΩ to few MΩ (**Figure 1f**). Representative device endurance with sub-ns reset pulses is shown in **Figure 1g** preserving a resistance ratio of ~100 within the first 1,500 switching cycles, and more ednurance data can be found in Figure S2. The measurements outlined below are performed within the first 1,000 cycles of each device, so no noticeable degradation is expected.

Our setup allows the measurement of transient PCM voltage $V_{PCM}(t)$ and current $I(t)$, hence transient power $P(t)$ and resistance $R_{trans}(t)$ are obtained directly, and the energy can be extracted by integrating the power over time. We define $V_{PG}(t)$ as the voltage applied by the pulse generator onto the 50 Ω output resistance, and $V_{scope}(t)$ as the voltage measured by the scope across the $Z_0$ = 50 Ω load. The current is $I(t) = V_{scope}(t)/Z_0$ and the voltage across the PCM is $V_{PCM} = V_{PG} - I(t)\cdot(Z_0+R_s) - V_{scope}(\mathrm{t})$, where $R_s \approx 400\ \Omega$ is the serial resistance of the lead, a thin W line connecting the BE to the pad, measured separately for each via size in devices without GST.

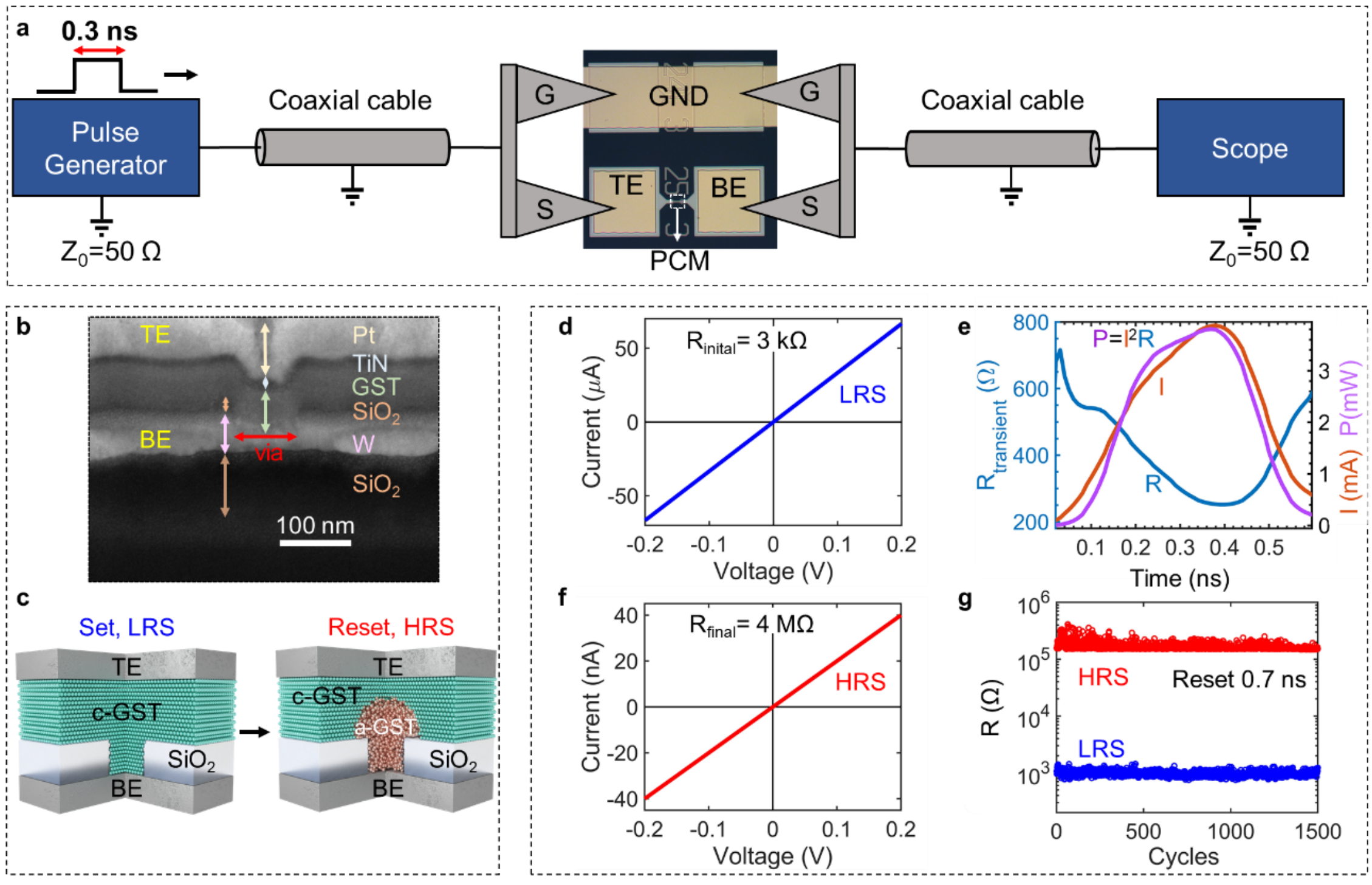

**Figure 1.** High-speed measurement setup and PCM device. (a) Schematic of the measurement setup. A pulse generator (PG) with output resistance of 50 Ω is connected by coaxial cable to an RF probe with the signal (S) to the top electrode (TE) pad of the PCM device and ground (G) to a local ground plane, shown in the optical image of the device. A fast scope is connected by RF probes (S) to the bottom electrode (BE) of the device. The RF probes are needed to minimize parasitic capacitance, and 50 Ω impedance matching suppresses signal reflections. (b) Cross section SEM image of a representative device. The phase change material is $Ge_2Sb_2Te_5$ (GST), sandwiched between Pt/TiN TE and W BE on a $SiO_2$/Si substrate. Sputtered $SiO_2$ serves as the insulator and the confined via hole diameter is marked by a horizontal red arrow. (c) Schematic cartoon illustrating the PCM device in its LRS or set state (crystalline, left), and the HRS, reset state, where the confined volume is amorphous (right). (d), (e), (f) Represent the measurements by their chronological order: (d) DC I-V sweep to read the initial resistance value in the set state, (e) reset pulse transient waveform, showing the evolution of the resistance ($V_{PCM}(t)/I_{PCM}(t)$), current ($I(t)$), and power ($I(t)\cdot V_{PCM}(t)$) during the pulse. The measured energy reported here is obtained by integrating $P(t)$ over time. The final resistance is read by another DC I-V sweep as shown in (f). Successful reset process results in HRS. (g) Endurance measurement of the first 1,500 cycles with sub-ns reset pulses. Details of the endurance program are found in the Methods section, and comparison with longer reset PW is shown in Figure S2.

## 2.2. Record-low reset energy density in PCM by sub-ns programming

Typically, PCM devices with 10 to 100 nm via diameter require 10 to 100 pJ for reset,[15,16] whereas ~0.1 pJ have been reported for PCM with carbon nanotube (CNT) electrodes thanks to the smaller volume to melt.[29] **Figure 2a,b** show a comparison between the reset energy for different PCM structures versus effective contact diameter. The purple markers represent results from the literature,[13,14,16–18,22,30–34] showing the scaling of energy with the effective diameter of the cell because less energy is needed to heat a smaller volume. We note that all previous studies from literature reporting reset energy used pulses wider than 3 ns, thus this has not been examined in the range of fast pulse widths pursued in this work.

Lines in Figure 2a,b represent calculated reset energy for a simple model of a GST sphere surrounded by $SiO_2$ (see Methods and Figure S3). The model roughly captures reset energy scaling with device dimensions, but more importantly it shows the energy can be reduced by orders of magnitude through a reduction of the PW. Energy dependence on PW is more significant for smaller devices because at narrower PW the heat has less time to diffuse outward and is better confined in the active PCM volume. For instance, the thermal diffusivity of $SiO_2$ is $\alpha_{ox} \sim 0.01$ cm$^2$s$^{-1}$, so the heat propagates a diffusion length $L_{th} \sim 1$ nm in 1 ps, ~10 nm in 0.1 ns, and ~300 nm in 100 ns, where $L_{th} = (\alpha t)^{1/2}$. Thus, for any PW longer than even a fraction of a nanosecond, the thermal diffusion length is greater than the device diameter, suggesting that a larger volume (of *e.g.*, dielectric) is heated *outside* the PCM cell than inside it.

It should also be noted that the shortest pulse to induce reset phase transition (by melt-quench) for a given cell size can be evaluated based on the speed of sound in the material ($v_s \sim 1$ nm/ps). In other words, based on this speed limit, the shortest reset pulse for a ~50 nm PCM cell would be ~50 ps. The adiabatic limit (solid black lines in Figure 2) is the energy to heat and melt the GST volume (including its latent heat of melting), without heating up its surrounding, *i.e.* assuming the GST is perfectly thermally isolated. For reference, the black dash-dot line (dashed in **Figure 2c**) represents the energy required to charge a pair of interconnect lines in order to program a single bit in a 1k×1k crossbar array.[15]

The measured reset energies of our PCM devices are also benchmarked against alternative emerging non-volatile memory (NVM) devices, namely spin torque transfer magnetic RAM (STT-MRAM), conductive bridge RAM (CBRAM) and resistive RAM (RRAM) in Figure 2c.[15] Our results show that reset energies within the same range of the PCM contacted by carbon nanotube (CNT, diameter ~1-2 nm) can be achieved in devices with more than 100× larger contact area (via diameter ~50 nm) by reducing the PW. For example, reset energy of ~0.2 pJ at PW of 0.3 ns was measured for ~50 nm via, and ~0.55 pJ at PW of 0.3 ns was measured for ~75 nm via. The comparison to other NVM devices in Figure 2c shows that the PCM energy achieved with sub-ns PW is comparable to energy consumption in alternative technologies and approaches the interconnect charging energy, a practical lower-limit target.

**Table 1** summarizes key results from the literature of reset energy consumption as well as sub-ns switching. Although several studies have shown sub-ns reset, the switching energy was not probed,[24,26,28,35–37] due to the challenge in measuring the current and voltage dynamics at such short timescale. Similarly, other studies have demonstrated reduction in reset energy by size scaling,[22,29,32] and improved thermal barriers,[17] but have not measured the switching energy for sub-ns pulses. Our results show that the reset energy is reduced as PW is decreased, and it should be emphasized that this reduction is in addition to other improvements in the energy consumption such as size scaling and thermal barrier engineering. Therefore the projected energy consumption for a PCM cell with 10 nm feature size at sub-ns reset pulse would be below 10 fJ.

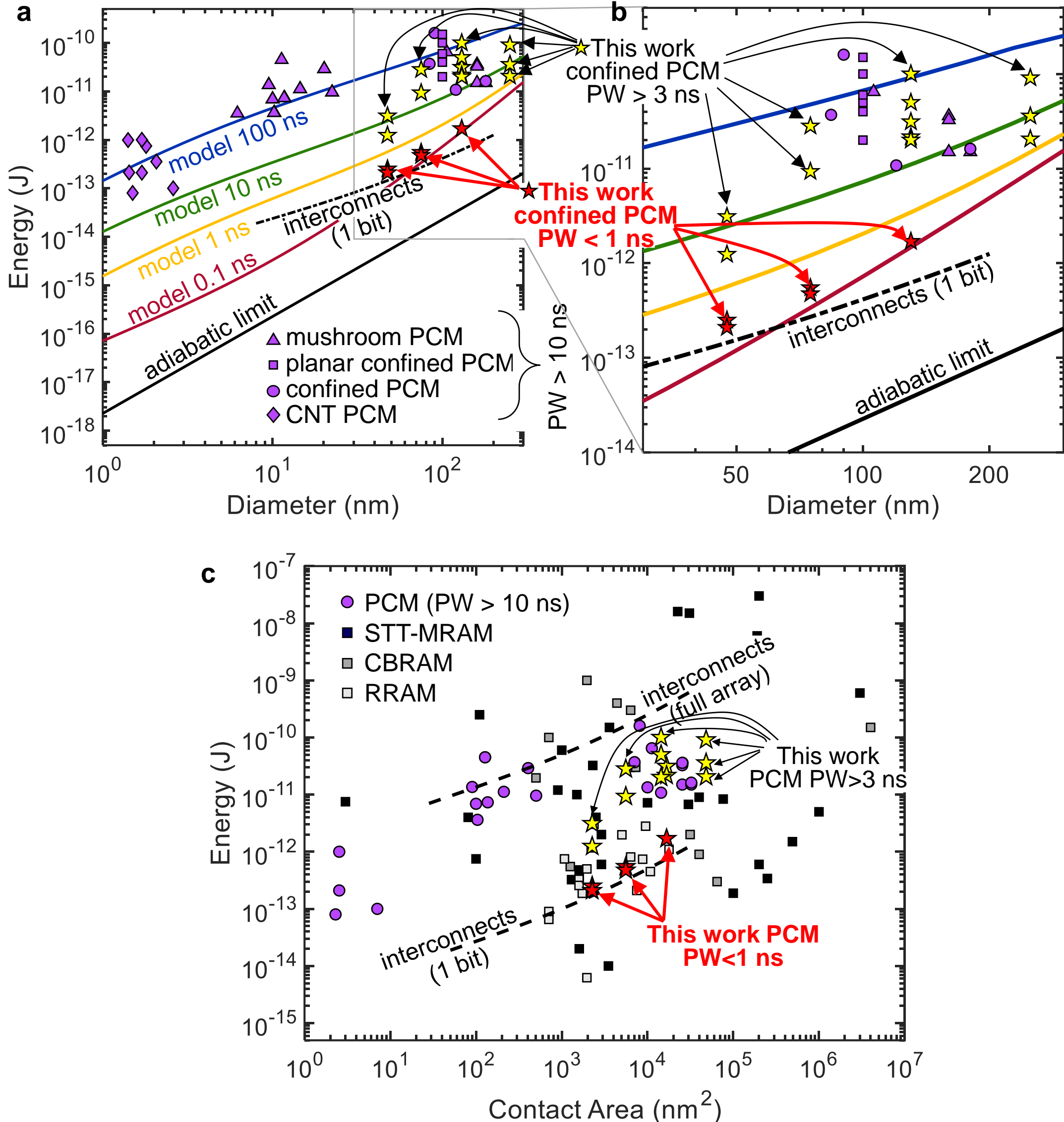

**Figure 2.** Benchmarking memory programming energy. (a),(b) PCM reset energy *versus* effective (confinement) diameter. Markers represent experimental data, lines represent calculations. All literature data of PCM reset energy consumption (purple markers) is for pulse width (PW) longer than 10 ns, whereas in this work we report on sub-ns reset programming (red stars) in addition to few ns to tens of ns pulses (yellow stars). The model lines represent calculated heating energy for GST sphere at $T_m$ surrounded by $SiO_2$ with varying pulse width. The adiabatic limit assumes perfectly isolated PCM, and the interconnects energy (dashed line) is the energy to charge only the metal lines for a single bit in a 1k × 1k crossbar array. (c) Programming energy of different emerging non-volatile memory devices, including PCM (same as in (a)), spin-torque transfer magnetic RAM (STT-MRAM), conductive-bridge RAM (CBRAM), and resistive RAM (RRAM). Lower dashed line (interconnects) is same as in (a), and the upper one is for a full 1k × 1k array.

**Table 1.** Comparison of state-of-the-art PCM reset energy and pulse width measurements.

| Device | PW (ns) | Energy (pJ) | Energy density (fJ/nm$^2$) | Size (nm) | Transient waveform | Comments and Ref. |
|---|---|---|---|---|---|---|
| GST mushroom | 100 | 15 | 0.46 | 180 | - | TaOx thermal barrier.[17] |
| GST in CNT crossbar | 50 | 1 | 345 | 1.7 | - | [32] |
| GST in CNT nano-gap | 20 | 0.1 | 14 | 2.6 | - | [22] |
| GST in CNT crossbar | 10 | 0.2 | 70 | 1.7 | - | [32] |
| GeTe mushroom | 2.5 [a)] | 7.5 | 2 | 60 | Shown | [25] |
| SST mushroom | 0.7 | - | - | 190 | Shown for V | Study focused at reducing set time.[36] |
| GST confined | 0.5 | - | - | 30 | Shown for V | Study focused at reducing set time.[24] |
| GST crossbar | 0.4 | - | - | - | - | [28] |
| GST confined | 0.4 | - | - | 20 | - | [35] |
| GST confined | 0.2 | - | - | 1000 | - | [26] |
| GST confined | 0.8 | 3.45 | 0.35 | 100 | Shown | This work |
| GST confined | 0.4 | 0.2 | 0.08 | 50 | Shown | This work |

[a)] Overall PW: rise+platue+fall = 2+1+2 ns

### 2.3. PCM reset energy limits and power-speed trade-off

PCM reset energy measurements show a clear benefit for PW reduction down to 0.3 ns, for devices with via diameter of 50 to 200 nm. Next, we turn to explore in more details the power-speed trade-off and energy limits of the reset process. **Figure 3** shows the final resistance following a reset pulse *vs*. (a) applied power and (b) current with varying PWs for a 75 nm via device. Successful reset is defined here for $R_{\mathrm{final}} > 100$ kΩ, *i.e.*, a change of nearly two orders of magnitude in resistance. Lower (higher) power and energy are required for smaller (larger) change in resistance, as shown in Figure S4. We observe that for PWs from 40 ns down to 6 ns, there is no appreciable difference in the reset power.[38] This is explained by the fact that the GST volume required to reset is already melted at <5 ns, and applying wider pulses mostly heats up the surrounding of the cell. It follows that for reset pulses longer than the thermal time constant of the GST region, reducing the PW will reduce the reset energy by the same factor because the power remains unchanged. Below PW of ~2 ns the reset power increases with the reduction of the PW. Amorphization requires ~3× to 4× higher power for 0.3 ns pulses compared with 6 ns (or longer). Although more power is needed, in terms of energy the 0.3 ns PW is preferred (~5× lower reset energy). Measured and simulated power-energy curves are shown in Figure S5.

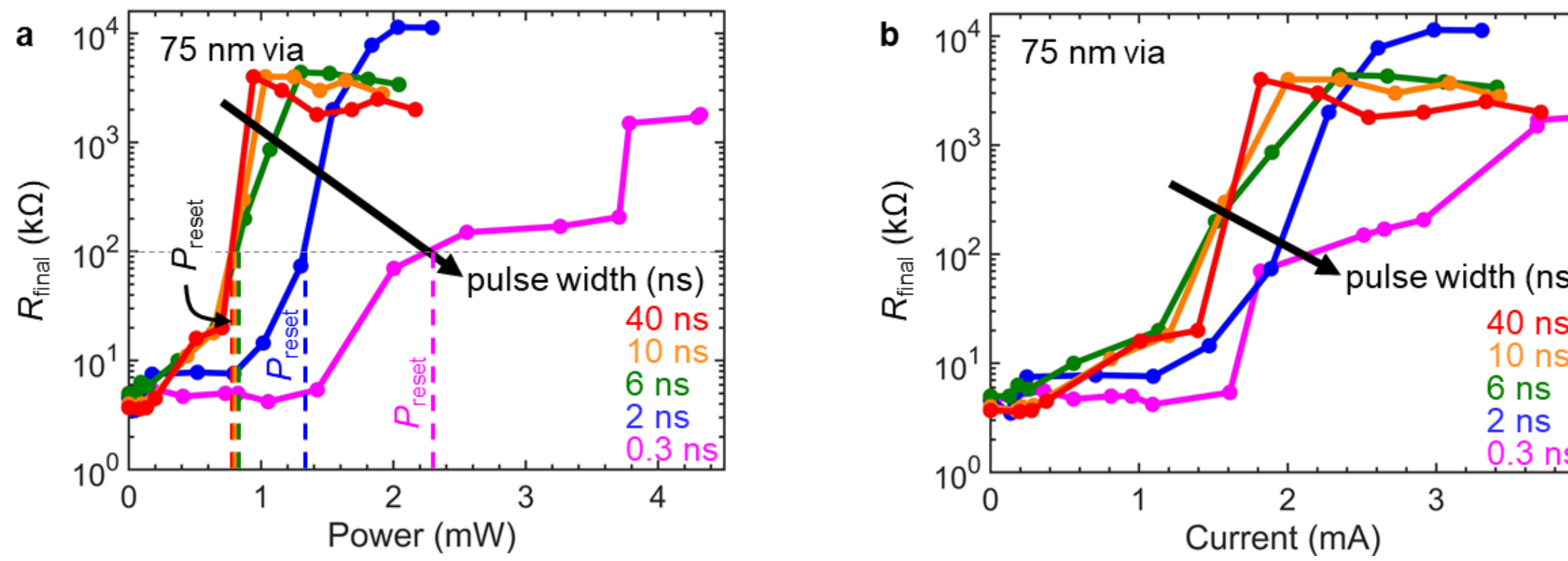

**Figure 3.** PCM reset power and current with varying pulse width. (a) Final read resistance versus maximum measured power (during reset pulse) with varying pulse width (PW) in the range 0.3 to 40 ns for confined PCM with ~75 nm via. The reset power remains unchanged for pulses longer than ~ 5 ns and starts to increase as the PW is reduced below ~3-4 ns. The reset powers (defined here for $R_{final}$ ~ 100 kΩ) are indicated by vertical dashed lines. (b) Final read resistance, same as (a) versus current. The initial resistance is in the range of 3-5 kΩ. Rise and fall times are set to 70 ps.

To better understand the relation between reset power and PW, we simulated the confined cells using FEM (see more details in Methods and Figure S6) for different via diameters and varying PW. **Figure 4a,b** show the measured and simulated reset energy versus PW for varying via sizes. Reset energy scales with via size as expected. Similarly, reset energy scales with PW down to ~1-2 ns, and the reset power remains constant in that range, as shown in **Figure 4c,d**. Below a critical PW the power increases. This critical time is shorter for smaller devices because it is proportional to the thermal time constant of the confined GST volume $\tau_{th}$ (where $\tau_{th}$ scales with device size).

This critical power-time point is predicted by our simulations and can be explained by the device temperature distribution (**Figure 4e**). Applying the minimal power for reset, the GST volume is melted within the thermal time constant of the confined GST, for example ~2-3 ns for ~100 nm via. For PW > $\tau_{th}$ (*e.g.* 30 ns) the only difference in the temperature map (compared with short PW ~ $\tau_{th}$) is further heating the surrounding of the cell (see Figure S7). However, applying the same power with PW < $\tau_{th}$ (*e.g.* 0.3 ns) the GST is not melted, and reset is not achieved. At such short PW, larger power is needed to reach the melting temperature ($T_m$) and achieve reset.

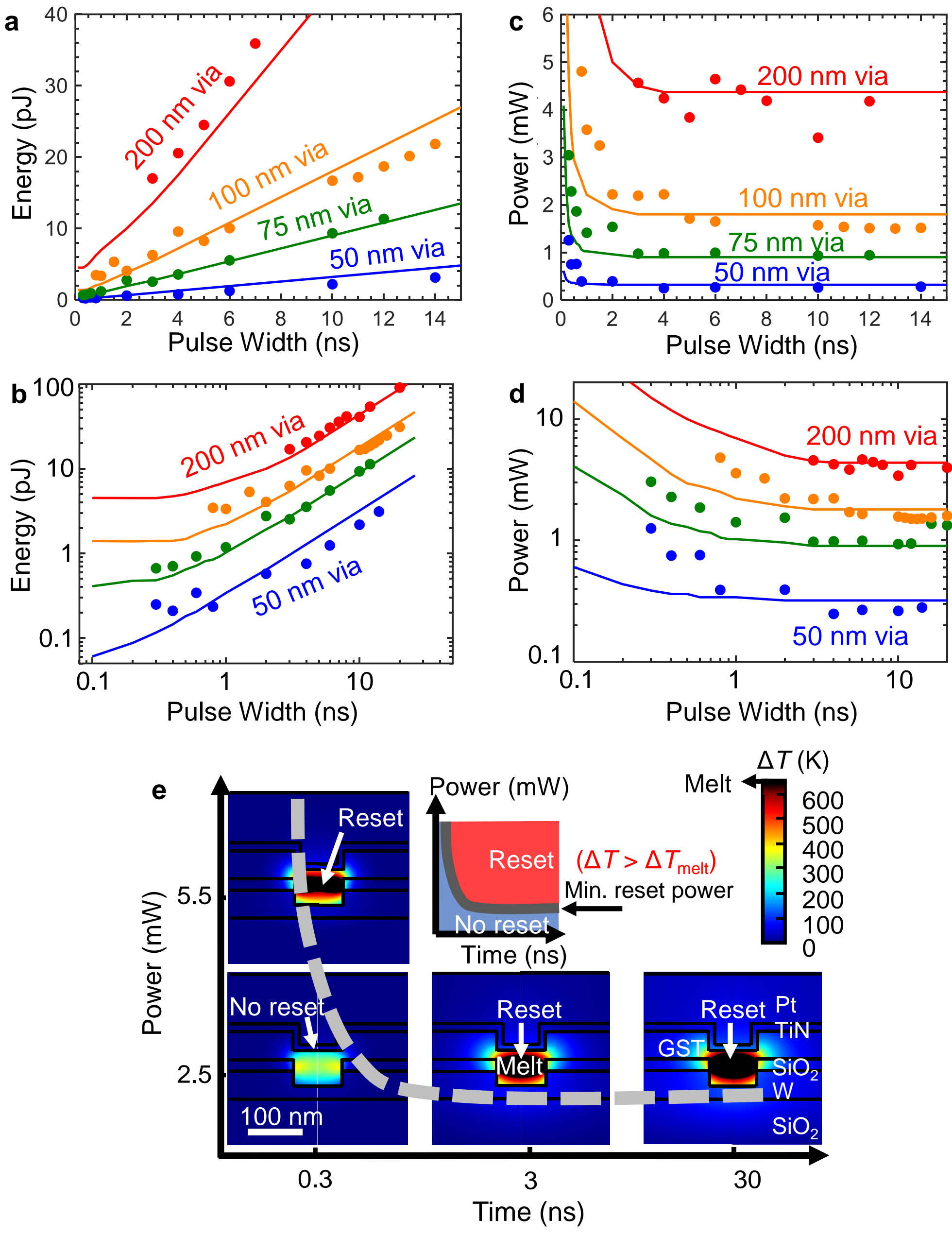

**Figure 4.** Switching energy and power-speed trade-off. (a) Linear and (b) log-log measured (markers) and simulated (lines) reset energy versus pulse width (PW) for devices with different via size (nominal ~ 50, 75, 100, 200 nm). (c) Linear and (d) log-log measured (markers) and simulated (lines) reset power versus PW for devices with different via size (nominal 50, 75, 100, 200 nm). (e) Maps of the temperature rise ($\Delta T$) obtained by finite element electro-thermal simulations of the PCM devices used in this work, arranged on power *versus* time (log-scale) with a schematic description of the dependence of minimum switching power vs time (or pulse width). In the temperature maps, black corresponds to melted volume ($\Delta T > 600$ K). The trend shows that the minimum reset power is nearly constant for PW longer than few ns and starts to increase as the PW is reduced below ~1-2 ns (roughly the thermal transient of the confined GST).

**2.4. PCM thermal dynamics**

The thermal time constant is a critical property of the device that determines the optimal programming PW conditions as outlined above. Besides simulations, an experimental approach is needed to evaluate the thermal dynamics of PCM. Previous work has shown that the resistivity of the GST in the liquid phase is smaller than in the crystalline phase (including high-temperature hexagonal GST).[39] Thus, probing the transient resistance $R_{trans}$ while the programming pulse is applied (Figure 1e) allows us to evaluate the thermal dynamics. During pulse application, $R_{trans}$ starts to decrease as the GST reaches $\sim\Delta T_m$ and starts to melt. $R_{trans}$ is further reduced as a larger volume of the GST is melted. Performing such measurements with sub-100 ps temporal resolution allows us to characterize the reset dynamics for sub-ns pulses. Transient measurements with via sizes of 100 nm, 75 nm, and 50 nm are shown in **Figure 5**. The transient voltage applied by the pulse generator ($V_{PG}$) and the voltage measured by the scope ($V_{scope}$) are shown in Figure 5a for a representative 75 nm via device. The transient resistance is extracted as $R_{trans} = V_{PCM}(t)/I(t)$ and plotted in Figure 5b with the current and power versus time. $R_{trans}$ scales inversely with via size (area), in agreement with a uniform (non-filamentary) melt-quench mechanism of PCM. The minimum measured values of $R_{trans}$ are ~440 Ω, ~200 Ω, and 130 Ω, for 50 nm, 75 nm, and 100 nm vias, respectively (Figure S8). Normalized $R_{trans}$ for different via sizes are depicted in Figure 5c, showing that the thermal transient is shorter for smaller devices. Previos work,[26,28] suggested non-thermal amorphization for sub-ns reset, but the transient resistance waveforms that we uncover here combined with electro-thermal simulations can explain the thermally induced melting and reset.

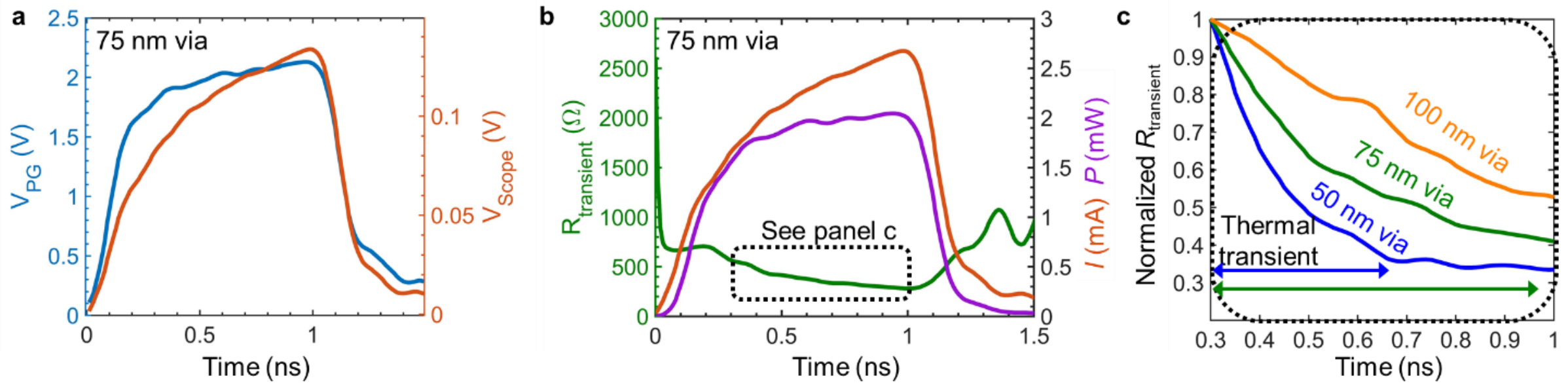

**Figure 5.** Dynamics of PCM reset. (a) Measured transient waveform of the voltage applied by the pulse generator ($V_{PG}$, blue) and the voltage probed by the scope (orange) during a 1 ns pulse for 75 nm via device. (b) Transient current ($I(t) = V_{scope}(t)/50\ \Omega$), power ($P(t) = V_{PCM}(t)\cdot I(t)$), and resistance ($R_{transient} = V_{PCM}(t)/I(t)$) during the 1 ns waveform. The voltage across the PCM is obtained by $V_{PCM} = V_{PG} - V_{scope} - I(t)\cdot(Z_0+R_S)$. (c) Normalized transient resistance versus time for devices with different via size showing the thermal transient of the GST confinement.

We point out that as device dimensions are reduced, the interfaces and contacts become more dominant both electrically and thermally.[40,41] Nonetheless, scaling down device dimensions is expected to effectively reduce the thermal time constant because the thermal capacitance keeps decreasing with PCM size. Finally, we note that for a given cell size, the final resistance is larger for lower $R_{trans}$ (during pulse application), since larger melt volume during the pulse (low $R_{trans}$) results in larger amorphous volume after the pulse (larger $R_{final}$).

## 3. CONCLUSIONS

We have reported nanosecond and sub-nanosecond pulsing and probing in confined PCM cells, showing record-low reset switching energy density (*e.g.*, less than 0.1 nJ/μm$^2$ at PW = 0.3 ns). We show that the reset power remains almost constant for pulses wider than the GST thermal time constant $\tau_{th}$. Extending the PW longer than $\tau_{th}$ (~1-2 ns) causes wasted energy by heating the dielectric surrounding the GST volume. For PW below $\tau_{th}$, higher power is needed to achieve reset, yet the overall reset energy is reduced within the shortest PW probed in our measurements (0.3 ns). Hence, constrained power systems should work with reset programming pulses on the order of ~ $\tau_{th}$, whereas high-performance computation systems could prefer lower write latencies at the expense of higher power consumption. Constrained energy systems could work with PW shorter than $\tau_{th}$ (sub-ns) if they can tolerate larger power consumption, *i.e*. larger instantaneous current or voltage. Finally, we also showed that the thermal dynamics of PCM can be probed by measuring the transient resistance of the device during pulse application. Our results shed new light on the energy limits of PCM operation and provide important guidelines for energy-efficient programming. Importantly, in the sub-ns regime, PCM reset energy approaches the charging energy of the interconnect lines in a crossbar array, a practical lower-limit benchmark for NVM and neuromorphic devices.

## METHODS

Device fabrication

The confined via-hole PCM is fabricated on a Si substrate with 100 nm thermally grown $SiO_2$. We start the process by defining the bottom electrode (BE). Tungsten (W) is evaporated, patterned with e-beam lithography, and dry-etched with reactive-ion etch (RIE). Next, we deposit $SiO_2$ using plasma-enhanced chemical vapor deposition (PECVD) and pattern the vias-hole using e-beam lithography. A 30-nm-thick $Ge_2Sb_2Te_5$ (GST) layer capped *in situ* with 20 nm TiN is DC sputtered and lifted-off. The GST/TiN layer is deposited *in situ* after the Ar sputter cleaning to prevent any native oxide formation. After GST/TiN lift-off, we pattern and lift-off an additional 20/40 nm sputtered TiN/Pt to form the probe pads and top electrode. Finally, a second metallization of 20/200 nm Ti/Au is e-beam evaporated and lifted-off, patterned with direct laser writer photolithography, to form the GS transmission line (TL).

Electrical measurements

The experimental setup (Figure 1a) consists of a fast pulse generator (PG) Active Technologies Pulse Rider PG-1072 connected in series with the PCM device and a Keysight Infiniium DSOS804A high-speed oscilloscope. The datasheet of the scope (https://www.keysight.com/us/en/assets/7018-04261/data-sheets/5991-3904.pdf) specifies typical rise/fall time of: 10/90%: 53.8 ps, 20/80%: 33.8 ps, and a temporal resolution of 3.15 ps. To support sub-nanosecond pulses, ground-signal (GS) transmission lines with characteristic impedance $Z_0$=50 Ω were used to connect the device with the measurement equipment. Transmission lines reduce pulse broadening, signal reflections, and possible ringing due to parasitic capacitances. The output impedance of the PG, as well as the input impedance of the scope are also set to 50 Ω. GGB GS RF probes are used in this setup, connected to the measurement equipment with 3.5 mm SMA cables.

Measurements were performed as follows. First, for each set of via sizes a "thru" configuration was measured, namely devices without GST are probed to obtain the transient waveform transmitted from the PG to the scope while applied to the cables, RF probes, device leads, without the PCM. Next, devices were programmed to LRS (~1 kΩ for 200-nm-via devices and ~10 kΩ for the 50-nm-via). Set pulses were 100 μs long with varying amplitude in the range ~1.4-2 V depending on the via size. The resistance of the cell was read with a DC bias voltage of 0.2 V.

Then, reset pulses with varying amplitude (0.8 to 5 V) and with varying pulse width (0.3 to 40 ns) were applied. Each reset pulse was followed by a read operation with a DC voltage of 0.2 V and a set pulse to return to the initial low resistance. The current through the PCM ($I(t)$) was calculated from the measured voltage at the scope divided by its input impedance ($Z_0 = 50\ \Omega$).

The endurance test was performed using Pulse Rider PG-1072 as pulse generator source and Keysight CX3322A Device Current Waveform Analyzer as current probe to read the resistance. Reset pulse widths were 0.7 ns, 5 ns, 10 ns, 40 ns with initial voltage amplitudes ranging from 6 V (for 0.7 ns PW) and between 3.6 and 4.8 V for the other PWs. Set PW was 5 μs with voltage amplitude of 2 V. The resistance was read by applying a 500 μs 0.2 V pulse and measuring the current using the CX3322A.

<u>FIB-SEM</u>

Cross-sectional images of the devices were made using focus-ion beam (FIB) scanning electron microscopy (SEM) in a Helios NanoLab G3 series DualBeam. The top-surface of the device is protected with 300-nm Pt sputtered in situ. A cross-section is milled using a 30 kV and 40 pA Ga beam. High resolution SEM micrographs were acquired at 5 kV and 10 nA beam with a 52-tilt angle using through lens detector (TLD) in topographical mode and in-column detector (ICD), which provides compositional contrast.

Simplified (spherical) reset energy model

The reset energy within our model is calculates as follows. We assume a GST sphere having diameter $D$ embedded in $SiO_2$ matrix. First, the adiabatic limit is calculated for which the GST sphere is perfectly isolated. The reset energy in this case is $E_{ad} = (C_s \cdot \Delta T_m + H) \cdot Vol$, where $C_s$ is the heat capacity of the GST, $\Delta T_m$ is the temperature rise required to reach the melting temperature, $H$ is the latent heat of melting, and $Vol$ is the GST volume. The other heating energies (with pulse width dependence) are obtained by first calculating the temperature rise distribution $\Delta T(r)$ for a given time $t$ (corresponding to the pulse width) while the GST sphere is at $\Delta T = \Delta T_m$, followed by integrating $C_s \cdot \Delta T(r)$ over the entire space. A thermal boundary resistance (TBR) is included by adding its equivalent thermal resistance to the calculation of the temperature distribution. A relatively low value of the TBR is used ~1 $m^2$K/GW, compared with measured TBR values for GST with its interfaces ($SiO_2$, TiN, W) of ~20-30 $m^2$K/GW for two reasons: 1) the relevant TBR for the model is at high-temperature (near the melting temperature), whereas the measured values are around room temperature, and 2) our simplified model assumes GST embedded in $SiO_2$, whereas real devices are contacted by metal electrodes which have thermal conductivity more than 10× larger compared with insulators such as $SiO_2$. Temperature maps of the model for varying PW are shown in Figure S3.

Finite element method electro-thermal simulations

Electro-thermal FEM simulations were performed in COMSOL, using a 2D axisymmetric finite element geometry. The bottom boundary of the Si substrate is held at ambient temperature, and the rest of the boundaries are insulating (adiabatic boundary condition). The BE is electrically grounded, and a voltage is applied to the TE. Device dimensions were taken from the FIB cross-sectional images (see Figure S1 and Table S1). The model includes electrothermal properties of the materials such as the temperature coefficient of resistance (TCR), the thermal conductivity ($k_{th}$), the heat capacity, and the thermal boundary resistance (TBR) at the interfaces.

Some of the key material parameters are outlined in Table S2. The GST electrical conductivity was taken from the melt value reported in ref, [42], the electrical contact resistance was determined by an effective thickness $t_{eff} = 10$ nm of the GST as defined in ref, [19]. The thermal properties of the PCM were taken from the upper boundary in ref [43] (because of the high temperature during reset). The TBR was chosen from the range of reported values in literature,[13,44] as a fitting parameter. More information on the distribution of voltage, current and power density is shown in Figure S5.

**Acknowledgements**

Fabrication of the confined PCM cells was performed at the Stanford Nanofabrication Facility (SNF) and Stanford Nano Shared Facilities (SNSF). Fabrication of the transmission lines was carried out at the Technion Micro-Nano Fabrication & Printing Unit (MNF&PU). We thank Dr. Larisa Popilevsky, RBNI FIB Lab. We thank Ido Kaminer, Dan Ritter, and Mario Lazna for careful reading of the manuscript. E.Y thanks Kye Okabe for fruitful discussions. This work was supported in part by ISF grant # 1179/20, by Russel Berrie Nanotechnology Institute (RBNI) seeding grant (NEVET), and by member companies of the Stanford Non-volatile Memory Technology Research Initiative (NMTRI). E.Y. is Northern Californian Career Development Chair Fellow.

**Supporting Information**

Cross sectional micrograph; Endurance test; Temperature maps in simplified spherical model; Power and energy consumption for different final resistance; Energy vs pulse width and power vs energy; Electro-thermal simulations; Effect of heat dissipation for $PW>\tau_{th}$; Transient resistance; Device dimensions from FIB-SEM micrograph; Material properties.

## REFERENCES

(1) Kim, T.; Lee, S. Evolution of Phase-Change Memory for the Storage-Class Memory and Beyond. *IEEE Trans. Electron Devices* **2020**, *67* (4), 1394–1406. https://doi.org/10.1109/TED.2020.2964640.

(2) Fong, S. W.; Neumann, C. M.; Wong, H. P. Phase-Change Memory — Towards a Storage-Class Memory. *IEEE Trans. Electron Devices* **2017**, *64* (11), 4374–4385.

(3) Burr, G. W.; Shelby, R. M.; Sebastian, A.; Kim, S.; Kim, S.; Sidler, S.; Virwani, K.; Ishii, M.; Narayanan, P.; Fumarola, A.; Sanches, L. L.; Boybat, I.; Le Gallo, M.; Moon, K.; Woo, J.; Hwang, H.; Leblebici, Y. Neuromorphic Computing Using Non-Volatile Memory. *Adv. Phys. X* **2017**, *2* (1), 89–124. https://doi.org/10.1080/23746149.2016.1259585.

(4) Yalon, E.; Datye, I. M.; Moon, J. S.; Son, K. A.; Lee, K.; Pop, E. Energy-Efficient Indirectly Heated Phase Change RF Switch. *IEEE Electron Device Lett.* **2019**, *40* (3), 455–458. https://doi.org/10.1109/LED.2019.2896953.

(5) Wainstein, N.; Adam, G.; Yalon, E.; Kvatinsky, S. Radio Frequency Switches Based on Emerging Resistive Memory Technologies: A Survey. *Proc. IEEE* **2020**, 1–19. https://doi.org/10.1109/jproc.2020.3011953.

(6) Ielmini, D.; Wong, H. S. P. In-Memory Computing with Resistive Switching Devices. *Nature Electronics*. Nature Publishing Group June 1, 2018, pp 333–343. https://doi.org/10.1038/s41928-018-0092-2.

(7) Le Gallo, M.; Sebastian, A.; Mathis, R.; Manica, M.; Giefers, H.; Tuma, T.; Bekas, C.; Curioni, A.; Eleftheriou, E. Mixed-Precision in-Memory Computing. *Nat. Electron.* **2018**, *1* (4), 246–253. https://doi.org/10.1038/s41928-018-0054-8.

(8) Sebastian, A.; Le Gallo, M.; Khaddam-Aljameh, R.; Eleftheriou, E. Memory Devices and Applications for In-Memory Computing. *Nature Nanotechnology*. Nature Research July 1, 2020, pp 529–544. https://doi.org/10.1038/s41565-020-0655-z.

(9) De Sandre, G.; Bettini, L.; Pirola, A.; Marmonier, L.; Pasotti, M.; Borghi, M.; Mattavelli, P.; Zuliani, P.; Scotti, L.; Mastracchio, G.; Bedeschi, F.; Gastaldi, R.; Bez, R. A 4 Mb LV MOS-Selected Embedded Phase Change Memory in 90 Nm Standard CMOS Technology. *IEEE J. Solid-State Circuits* **2011**, *46* (1), 52–63. https://doi.org/10.1109/JSSC.2010.2084491.

(10) Intel 3D XPoint Memory Die Removed from Intel Optane$^{TM}$ PCM (Phase Change Memory) https://www.techinsights.com/blog/intel-3d-xpoint-memory-die-removed-intel-optanetm-pcm-phase-change-memory (accessed Aug 31, 2020).

(11) Ovshinsky, S. R. Reversible Electrical Switching Phenomena in Disordered Structures. *Phys. Rev. Lett.* **1968**, *21* (20), 1450–1453. https://doi.org/10.1103/PhysRevLett.21.1450.

(12) Raoux, S.; Xiong, F.; Wuttig, M.; Pop, E. Phase Change Materials and Phase Change Memory. *MRS Bull.* **2014**, *39* (8), 703–710. https://doi.org/10.1557/mrs.2014.139.

(13) Wong, H. S. P.; Raoux, S.; Kim, S.; Liang, J.; Reifenberg, J. P.; Rajendran, B.; Asheghi, M.; Goodson, K. E. Phase Change Memory. *Proc. IEEE* **2010**, *98* (12), 2201–2227. https://doi.org/10.1109/JPROC.2010.2070050.

(14) Xiong, F.; Yalon, E.; Behnam, A.; Neumann, C. M.; Grosse, K. L.; Deshmukh, S.; Pop, E. Towards Ultimate Scaling Limits of Phase-Change Memory. *IEEE Int. Electron Devices Meet.* **2016**, 79–82.

(15) H.-S. P. Wong, C. Ahn, J. Cao, H.-Y. Chen, S. B. Eryilmaz, S. W. Fong, J. A. Incorvia,

Z. Jiang, H. Li, C. Neumann, K. Okabe, S. Qin, J. Sohn, Y. Wu, S. Yu, X. Z. Stanford Memory Trends | Stanford Nanoelectronics Lab https://nano.stanford.edu/stanford-memory-trends/.

(16) Lacaita, A. L.; Redaelli, A. The Race of Phase Change Memories to Nanoscale Storage and Applications. *Microelectron. Eng.* **2013**, *109*, 351–356. https://doi.org/10.1016/j.mee.2013.02.105.

(17) Matsui, Y.; Kurotsuchi, K.; Tonomura, O.; Morikawa, T.; Kinoshita, M.; Fujisaki, Y.; Matsuzaki, N.; Hanzawa, S.; Terao, M.; Takaura, N.; Moriya, H.; Iwasaki, T.; Moniwa, M.; Koga, T. $Ta_2O_5$ Interfacial Layer between GST and W Plug Enabling Low Power Operation of Phase Change Memories. *Tech. Dig. - Int. Electron Devices Meet. IEDM* **2006**, 2–5. https://doi.org/10.1109/IEDM.2006.346908.

(18) Fong, S. W.; Neumann, C. M.; Yalon, E.; Rojo, M. M.; Pop, E.; Wong, H. S. P. Dual-Layer Dielectric Stack for Thermally Isolated Low-Energy Phase-Change Memory. *IEEE Trans. Electron Devices* **2017**, *64* (11), 4496–4502. https://doi.org/10.1109/TED.2017.2756071.

(19) Neumann, C. M.; Okabe, K. L.; Yalon, E.; Grady, R. W.; Wong, H.-S. P.; Pop, E. Engineering Thermal and Electrical Interface Properties of Phase Change Memory with Monolayer $MoS_2$. *Appl. Phys. Lett.* **2019**, *114*, 82103. https://doi.org/10.1063/1.5080959.

(20) Ahn, C.; Fong, S. W.; Kim, Y.; Lee, S.; Sood, A.; Neumann, C. M.; Asheghi, M.; Goodson, K. E.; Pop, E.; Wong, H. S. P. Energy-Efficient Phase-Change Memory with Graphene as a Thermal Barrier. *Nano Lett.* **2015**, *15* (10), 6809–6814. https://doi.org/10.1021/acs.nanolett.5b02661.

(21) Im, D. H.; Lee, J. I.; Cho, S. L.; An, H. G.; Kim, D. H.; Kim, I. S.; Park, H.; Ahn, D. H.; Horii, H.; Park, S. O.; Chung, U. I.; Moon, J. T. A Unified 7.5nm Dash-Type Confined Cell for High Performance PRAM Device. *Tech. Dig. - Int. Electron Devices Meet. IEDM* **2008**. https://doi.org/10.1109/IEDM.2008.4796654.

(22) Xiong, F.; Liao, A. D.; Estrada, D.; Pop, E. Low-Power Switching of Phase-Change Materials with Carbon Nanotube Electrodes. *Science (80-. ).* **2011**, *332* (6029), 568–570. https://doi.org/10.1126/science.1201938.

(23) Pop, E. Energy Dissipation and Transport in Nanoscale Devices. *Nano Res.* **2010**, *3* (3), 147–169. https://doi.org/10.1007/s12274-010-1019-z.

(24) Loke, D.; Lee, T. H.; Wang, W. J.; Shi, L. P.; Zhao, R.; Yeo, Y. C.; Chong, T. C.; Elliott, S. R. Breaking the Speed Limits of Phase-Change Memory. *Science (80-. ).* **2012**, *336* (6088), 1566–1569.

(25) Bruns, G.; Merkelbach, P.; Schlockermann, C.; Salinga, M.; Wuttig, M.; Happ, T. D.; Philipp, J. B.; Kund, M. Nanosecond Switching in GeTe Phase Change Memory Cells. *Appl. Phys. Lett.* **2009**, *95* (4), 93–96. https://doi.org/10.1063/1.3191670.

(26) Huang, D. Q.; Miao, X. S.; Li, Z.; Sheng, J. J.; Sun, J. J.; Peng, J. H.; Wang, J. H.; Chen, Y.; Long, X. M. Nonthermal Phase Transition in Phase Change Memory Cells Induced by Picosecond Electric Pulse. *Appl. Phys. Lett.* **2011**, *98* (24), 242106. https://doi.org/10.1063/1.3597792.

(27) Loke, D.; Skelton, J. M.; Wang, W. J.; Lee, T. H.; Zhao, R.; Chong, T. C.; Elliott, S. R. Ultrafast Phase-Change Logic Device Driven by Melting Processes. *Proc. Natl. Acad. Sci. U. S. A.* **2014**, *111* (37), 13272–13277. https://doi.org/10.1073/pnas.1407633111.

(28) Wang, P.; Ju, C.; Chen, W.; Huang, D. Q.; Guan, X. W.; Li, Z.; Cheng, X. M.; Miao, X. S. Picosecond Amorphization of Chalcogenides Material: From Scattering to

Ionization. *Appl. Phys. Lett.* **2013**, *102* (11), 112108. https://doi.org/10.1063/1.4798263.

(29) Xiong, F.; Bae, M. H.; Dai, Y.; Liao, A. D.; Behnam, A.; Carrion, E. A.; Hong, S.; Ielmini, D.; Pop, E. Self-Aligned Nanotube-Nanowire Phase Change Memory. *Nano Lett.* **2013**, *13* (2), 464–469. https://doi.org/10.1021/nl3038097.

(30) Oh, H. R.; Cho, B. H.; Cho, W. Y.; Kang, S.; Choi, B. G.; Kim, H. J.; Kim, K. S.; Kim, D. E.; Kwak, C. K.; Byun, H. G.; Jeong, G. T.; Jeong, H. S.; Kim, K. Enhanced Write Performance of a 64-Mb Phase-Change Random Access Memory. *IEEE J. Solid-State Circuits* **2006**, *41* (1), 122–126. https://doi.org/10.1109/JSSC.2005.859016.

(31) Annunziata, R.; Zuliani, P.; Borghi, M.; De Sandre, G.; Scotti, L.; Prelini, C.; Tosi, M.; Tortorelli, I.; Pellizzer, F. Phase Change Memory Technology for Embedded Non Volatile Memory Applications for 90nm and Beyond. *Tech. Dig. - Int. Electron Devices Meet. IEDM* **2009**, 97–100. https://doi.org/10.1109/IEDM.2009.5424413.

(32) Liang, J.; Jeyasingh, R. G. D.; Chen, H. Y.; Wong, H. S. P. An Ultra-Low Reset Current Cross-Point Phase Change Memory with Carbon Nanotube Electrodes. *IEEE Trans. Electron Devices* **2012**, *59* (4), 1155–1163. https://doi.org/10.1109/TED.2012.2184542.

(33) Evans, E. J.; Helbers, J. H.; Ovshinsky, S. R. Reversible Conductivity Transformations in Chalcogenide Alloy Films. *J. Non. Cryst. Solids* **1970**, *2* (C), 334–346. https://doi.org/10.1016/0022-3093(70)90149-3.

(34) Lee, S. H.; Park, H. C.; Kim, M. S.; Kim, H. W.; Choi, M. R.; Lee, H. G.; Seo, J. W.; Kim, S. C.; Kim, S. G.; Hong, S. B.; Lee, S. Y.; Lee, J. U.; Kim, Y. S.; Kim, K. S.; Kim, J. I.; Lee, M. Y.; Shin, H. S.; Chae, S. J.; Song, J. H.; Yoon, H. S.; Oh, J. M.; Min, S. K.; Lee, H. M.; Hong, K. R.; Cheong, J. T.; Park, S. N.; Ku, J. C.; Shin, H. S.; Sohn, Y. S.; Park, S. K.; Kim, T. S.; Kim, Y. K.; Park, K. W.; Han, C. S.; Kim, H. W.; Kim, W.; Kim, H. J.; Choi, K. S.; Lee, J. H.; Hong, S. J. Highly Productive PCRAM Technology Platform and Full Chip Operation: Based on $4F^2$ (84nm Pitch) Cell Scheme for 1 Gb and Beyond. *Tech. Dig. - Int. Electron Devices Meet. IEDM* **2011**, *2*, 47–50. https://doi.org/10.1109/IEDM.2011.6131480.

(35) Wang, W. J.; Shi, L. P.; Zhao, R.; Lim, K. G.; Lee, H. K.; Chong, T. C.; Wu, Y. H. Fast Phase Transitions Induced by Picosecond Electrical Pulses on Phase Change Memory Cells. *Appl. Phys. Lett.* **2008**, *93* (4), 43121. https://doi.org/10.1063/1.2963196.

(36) Rao, F.; Ding, K.; Zhou, Y.; Zheng, Y.; Xia, M.; Lv, S.; Song, Z.; Feng, S.; Ronneberger, I.; Mazzarello, R.; Zhang, W.; Ma, E. Reducing the Stochasticity of Crystal Nucleation to Enable Subnanosecond Memory Writing. *Science (80-. ).* **2017**, *358* (6369), 1423–1427. https://doi.org/10.1126/science.aao3212.

(37) Loke, D. K.; Skelton, J. M.; Lee, T. H.; Zhao, R.; Chong, T. C.; Elliott, S. R. Ultrafast Nanoscale Phase-Change Memory Enabled by Single-Pulse Conditioning. *ACS Appl. Mater. Interfaces* **2018**, *10* (49), 41855–41860. https://doi.org/10.1021/acsami.8b16033.

(38) Yalon, E.; Okabe, K.; Neumann, C. M.; Wong, H. S. P.; Pop, E. Energy-Efficient Phase Change Memory Programming by Nanosecond Pulses. *Device Res. Conf. - Conf. Dig. DRC* **2018**, *2018-June* (2016), 1–2. https://doi.org/10.1109/DRC.2018.8443164.

(39) Cil, K.; Member, S.; Dirisaglik, F.; Member, S.; Adnane, L.; Wennberg, M.; King, A.; Faraclas, A.; Akbulut, M. B.; Member, S.; Zhu, Y.; Lam, C.; Gokirmak, A.; Member, S.; Silva, H.; Member, S.; The, A.; Sb, G.; Gst, T. Electrical Resistivity of Liquid $Ge_2Sb_2Te_5$ Based on Thin-Film and Nanoscale Device Measurements. *IEEE Trans.*

*Electron Devices* **2013**, *60* (1), 433–437.

(40) Reifenberg, J. P.; Kencke, D. L.; Goodson, K. E. The Impact of Thermal Boundary Resistance in Phase-Change Memory Devices. *IEEE Electron Device Lett.* **2008**, *29* (10), 1112–1114. https://doi.org/10.1109/LED.2008.2003012.

(41) Kencke, D. L.; Karpov, I. V.; Johnson, B. G.; Lee, S. J.; Kau, D.; Hudgens, S. J.; Reifenberg, J. P.; Savransky, S. D.; Zhang, J.; Giles, M. D.; Spadini, G. The Role of Interfaces in Damascene Phase-Change Memory. *Tech. Dig. - Int. Electron Devices Meet. IEDM* **2007**, 323–326. https://doi.org/10.1109/IEDM.2007.4418936.

(42) N. Wainstein, G. Ankonina, S. Kvatinsky, and E. Y. Compact Modeling and Electro-Thermal Measurements of Indirectly-Heated Phase Change RF Switches. *IEEE Trans. Electron Devices* **2020**.

(43) Bozorg-Grayeli, E.; Reifenberg, J. P.; Asheghi, M.; Wong, H.-S. P.; Goodson, K. E. THERMAL TRANSPORT IN PHASE CHANGE MEMORY MATERIALS. *Annu. Rev. Heat Transf.* **2013**, *16* (1), 397–428. https://doi.org/10.1615/annualrevheattransfer.v16.130.

(44) Yalon, E.; Deshmukh, S.; Muñoz Rojo, M.; Lian, F.; Neumann, C. M.; Xiong, F.; Pop, E. Spatially Resolved Thermometry of Resistive Memory Devices. *Sci. Rep.* **2017**, *7* (1). https://doi.org/10.1038/s41598-017-14498-3.

Supporting Information

# Uncovering Phase Change Memory Energy Limits by Sub-Nanosecond Probing of Power Dissipation Dynamics

Keren Stern[1,†], Nicolás Wainstein[1,†], Yair Keller[1], Christopher M. Neumann[2], Eric Pop[2,3], Shahar Kvatinsky[1], and Eilam Yalon[1,*]

[1]*Electrical Engineering, Technion - Israel Institute of Technology, Haifa, 32000, Israel*

[2]*Electrical Engineering, Stanford University, Stanford, CA 94305, USA.*

[3]*Materials Science and Engineering, Stanford University, Stanford, CA 94305, USA.*

[†]These authors contributed equally

[*]Contact: eilamy@technion.ac.il

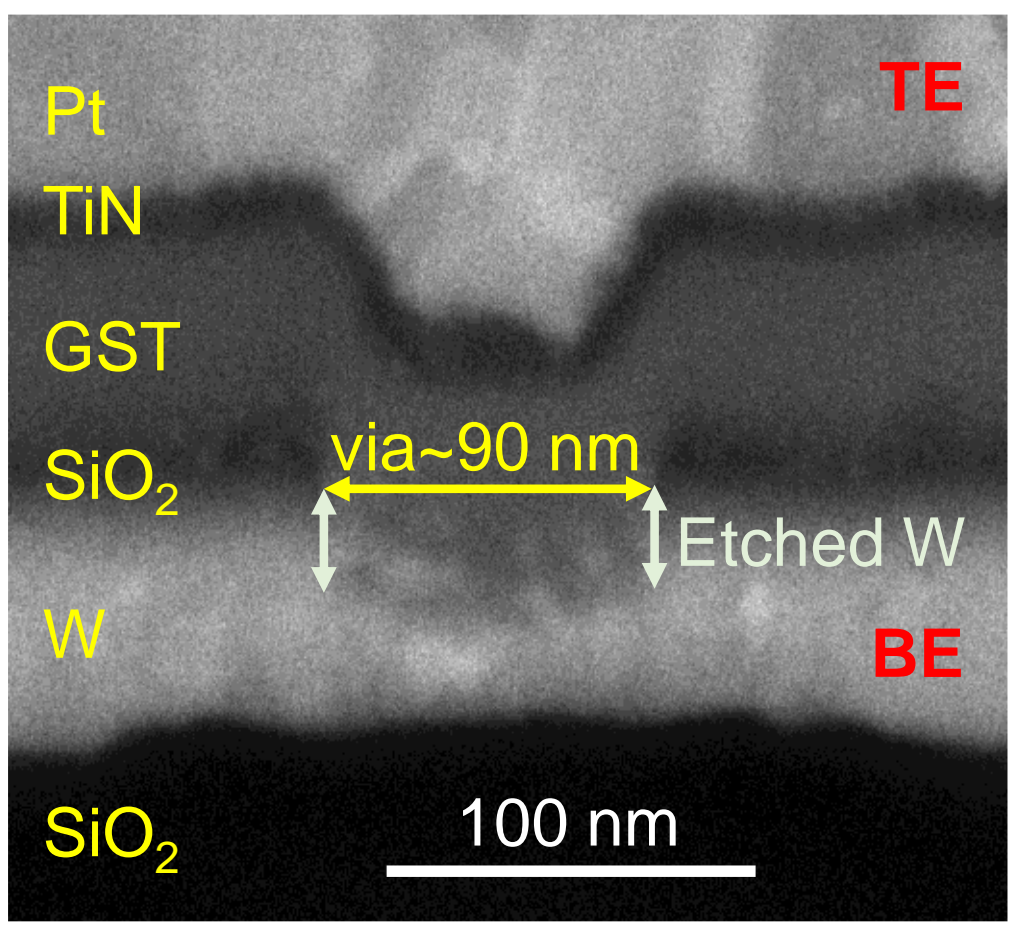

**Figure S1. Cross sectional micrograph.** Cross section was obtain using FIB and captured by SEM in column detector (ICD) that provides compositional contrast as described in the Methods section. The device shown is for nominal 100 nm via. The marked ~90 nm is the via diameter/length, but the GST-W contact (PCM-BE) consists also of the etched W "walls" (marked by vertical arrows). The overall contact area for each device was taken as the sum of the via diameter squared plus four rectangular areas of via length multiplied by the etched W height. The measured values for all devices (with varying nominal via size) are summarized in Table S1. These values were used in the finite element simulations.

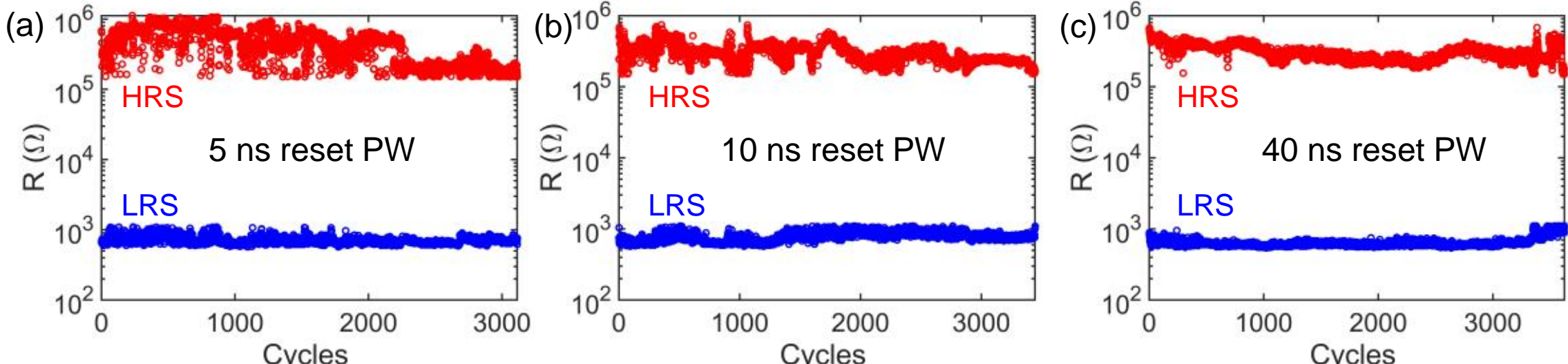

**Figure S2. Endurance test.** Endurance test with varying reset pulse width of (a) 5 ns, (b) 10 ns, and (c) 40 ns. The details of the endurance test are given in the Methods Section. Failure of our devices typically occurs after few thousand cycles. The switching is stable in the first ~1,000 cycles. The measurements reported in the main text are within the first 1,000 cycles and no degradation is expected.

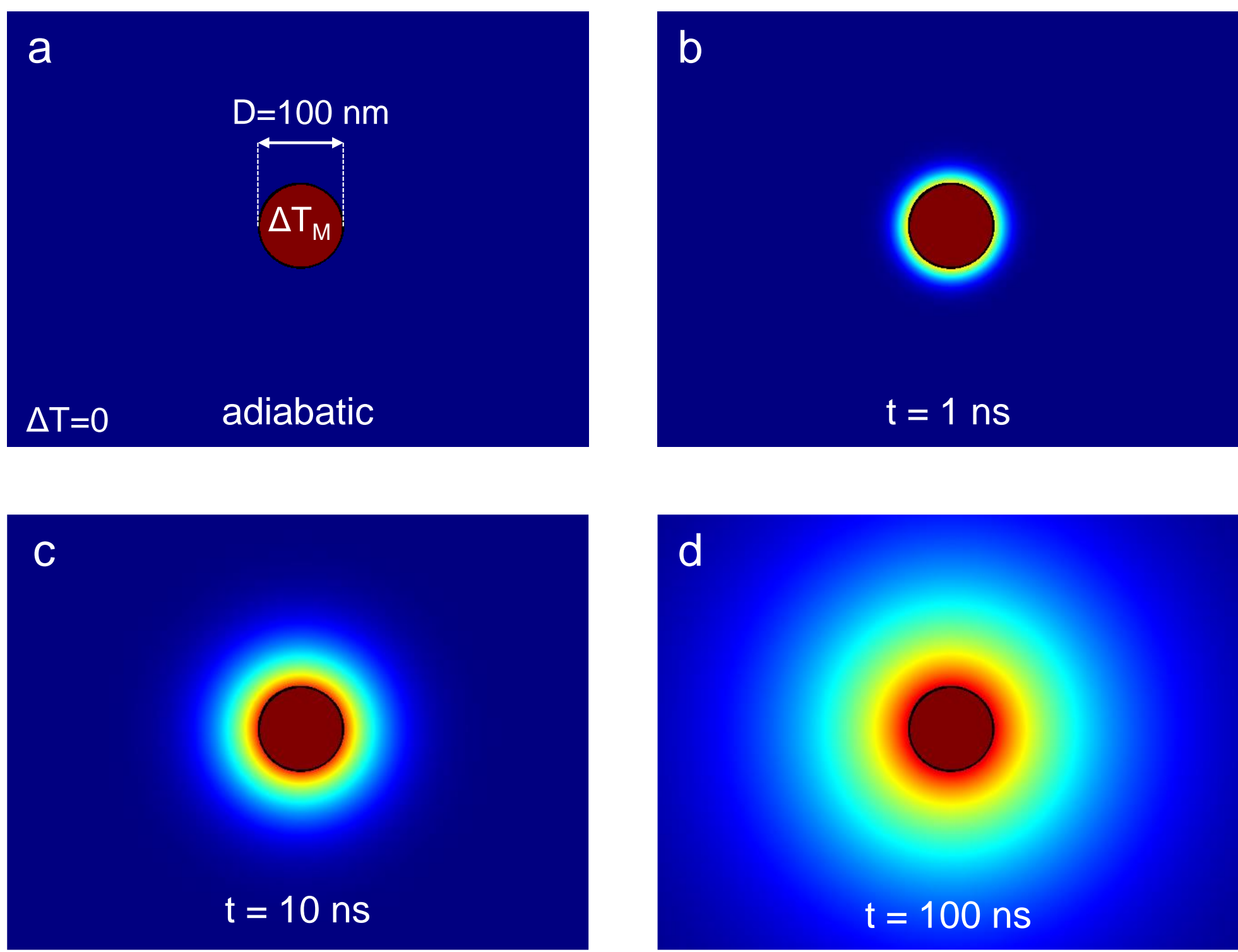

**Figure S3. Temperature maps in simplified spherical model**. Map of the temperature rise in our simplified model of GST sphere at $T$=$T_m$ in $SiO_2$ under different condition: (**a**) the adiabatic case (GST is perfectly isolated), (**b**) PW=1 ns, (**c**) PW=10 ns, and (**d**) PW=100 ns.

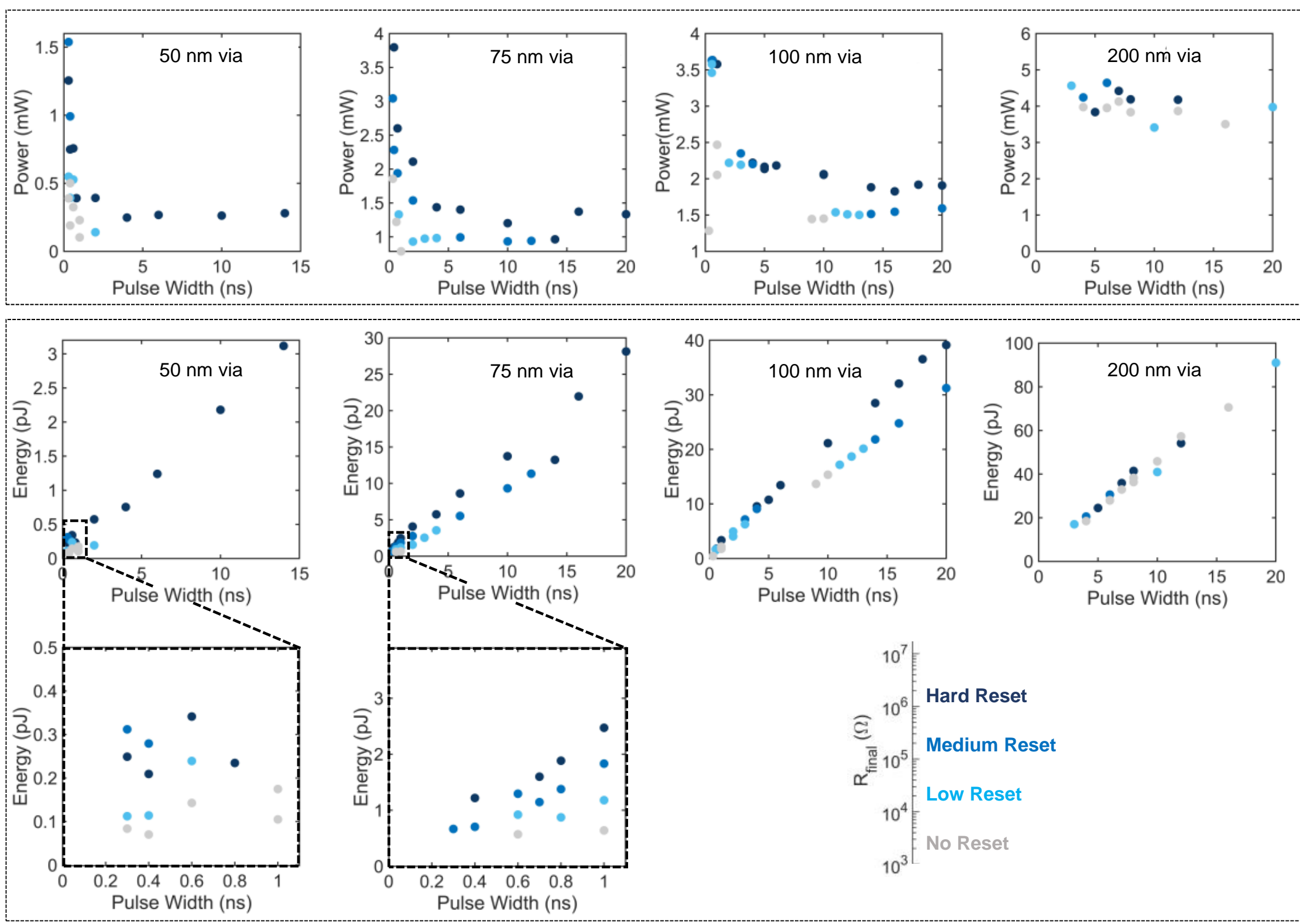

**Figure S4. Power and energy consumption for different final resistance**. Following the reset pulses, final resistances were obtained with different values across several orders of magnitude. In the main text reset is defined for $R_{\text{final}}$ > 100 kΩ. This figure shows power and energy consumption for varying pulse width; markers' color represent the final resistance, as follows: no reset (grey, $R_{\text{final}}$<10 kΩ), low reset (light blue, 10 kΩ <$R_{\text{final}}$<100 kΩ), medium reset (blue, 100 kΩ <$R_{\text{final}}$<1 MΩ), and hard reset (dark blue, $R_{\text{final}}$>1 MΩ).

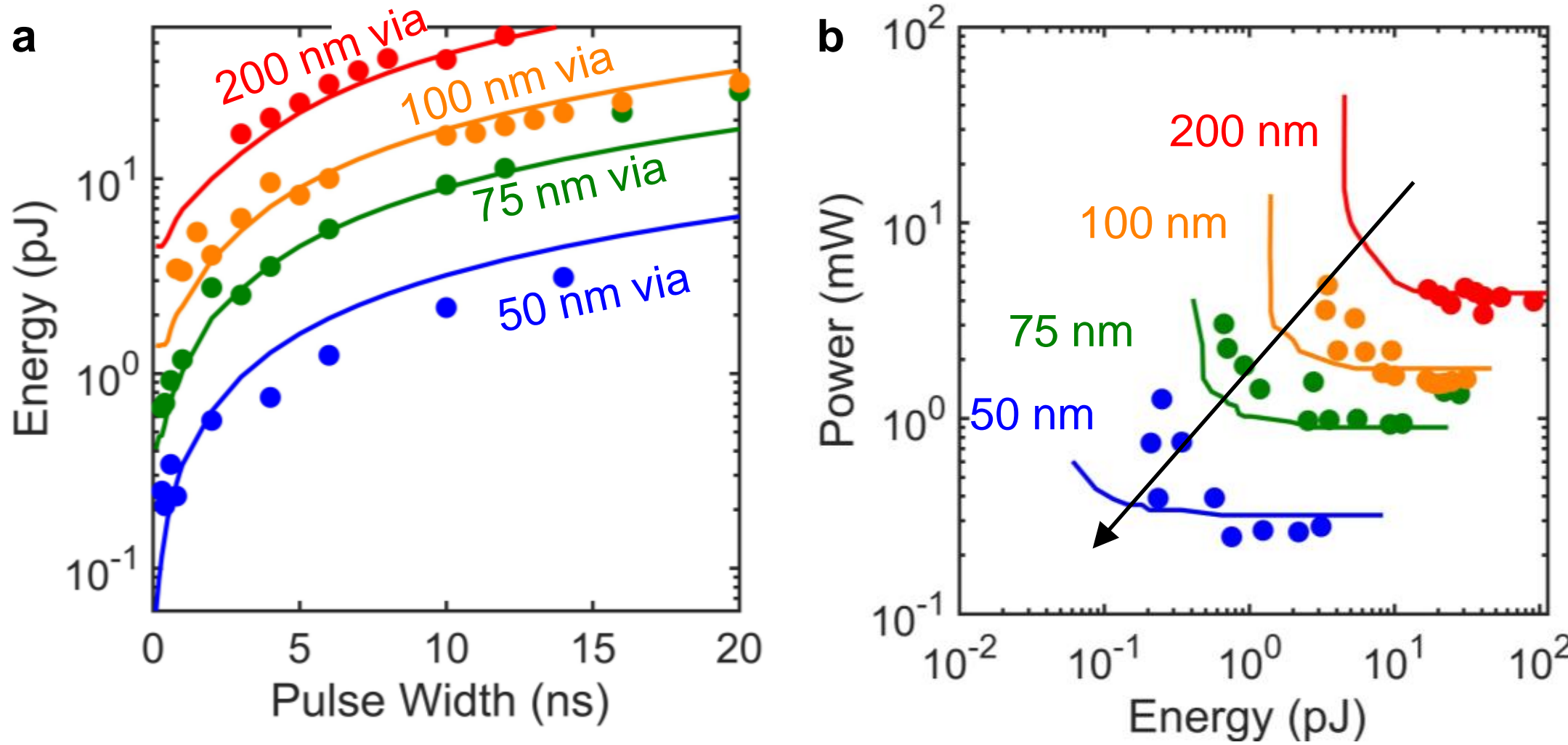

**Figure S5. Energy vs pulse width and power vs energy.** (**a**) Reset energy (log-scale) versus pulse width. (**b**) Log-scale plot of power versus energy. Markers represent experimental measurements and lines represent FEM simulations. Reset energies of sub-pJ are obtained for 50 and 75 nm via devices at sub-ns PW.

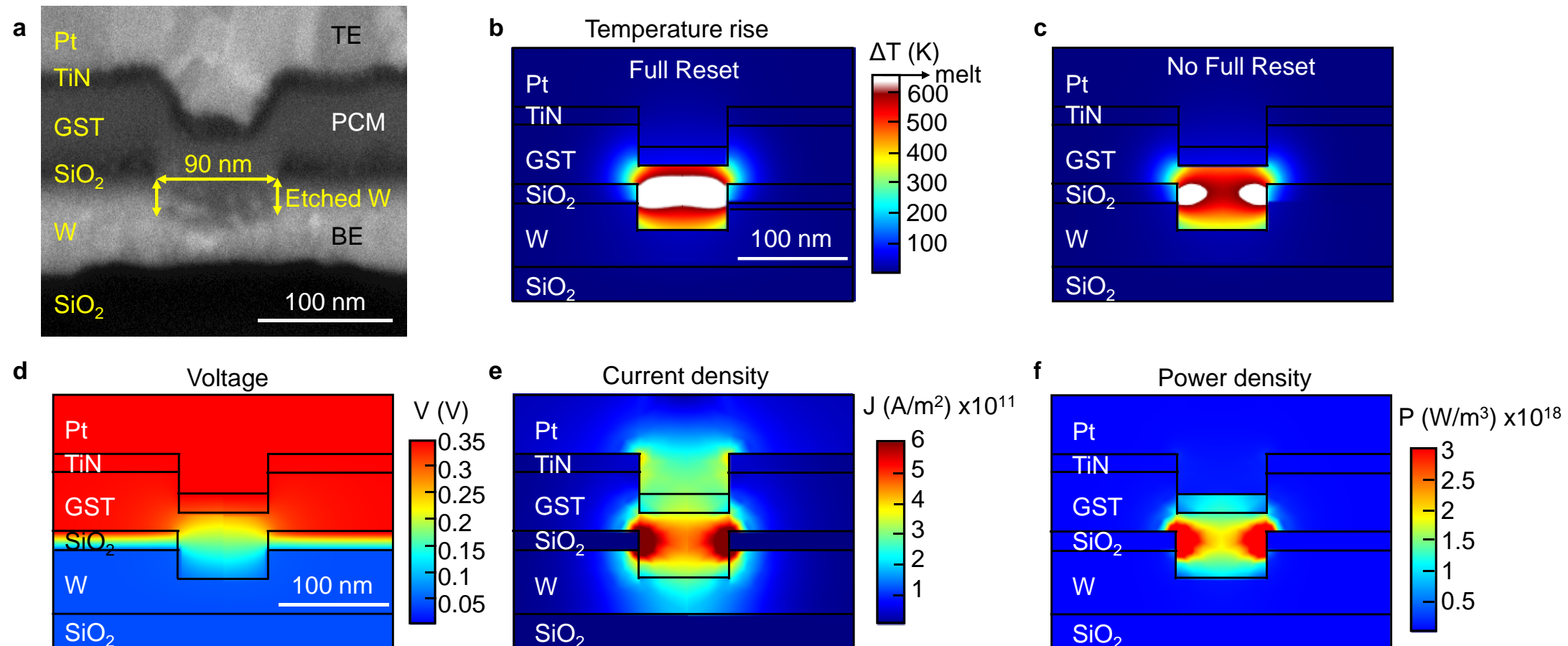

**Figure S6. Electro-thermal simulations.** **(a)** Cross-sectional micrograph of ~100 nm nominal via size, same as Supplementary Fig. 1 to show the dimensions of the device. (**b**) Temperature rise ($\Delta T$) map of 100 nm nominal via device at PW=0.4 ns at the conditions of full reset. (**c**) Same as (**b**) with only partial melting, where conditions of full reset are not met. Maps of (**d**) voltage, (**e**) current density, and (**f**) power density during full reset at PW=0.4 ns.

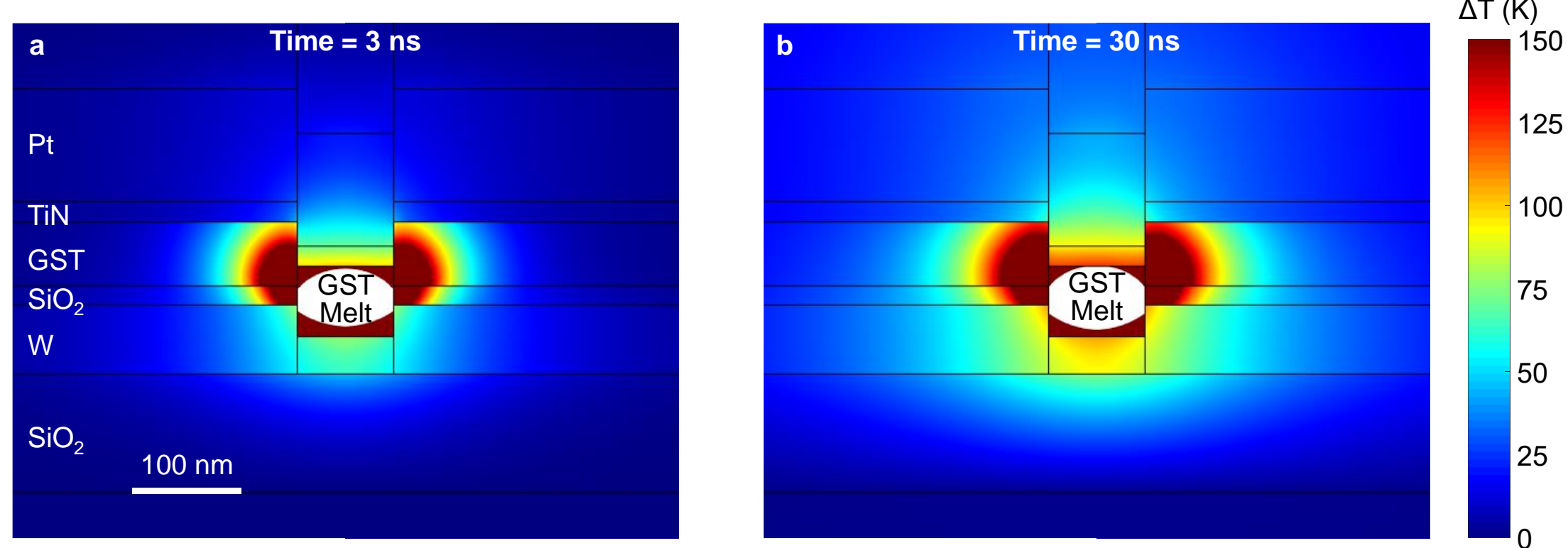

**Figure S7. Effect of heat dissipation for PW>$\tau_{th}$.** Comparison of the simulated heat dissipation in a nominal 100 nm via cell at **(a)** PW=3 ns (~$\tau_{th}$) and **(b)** PW=30 ns at the same conditions as in Fig. 4e in the main text ($P$=2.5 mW). The temperature rise ($\Delta T$) color bar scales only up to $\Delta T$=150 K to illustrate the heat wasted to the surrounding in the longer pulse of 30 ns, whereas reset (GST melt) is achieved in less than 3 ns.

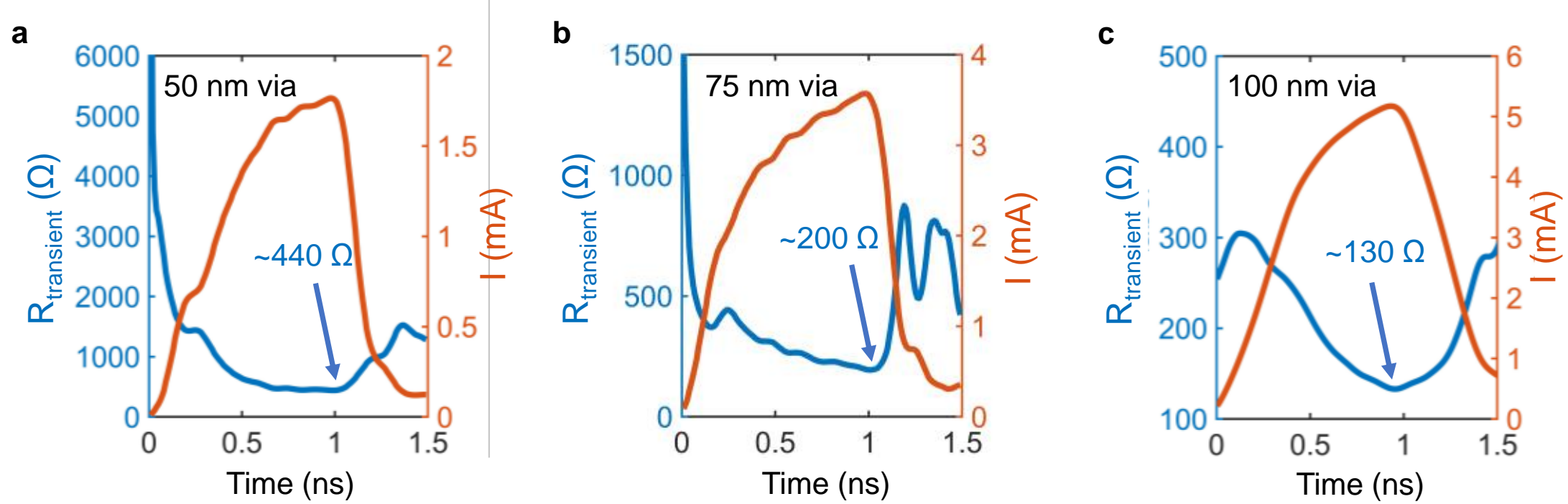

**Figure S8. Transient resistance.** Comparison of the transient waveform $R_{\text{transient}}$ and $I(t)$ for three different via sizes: (**a**) 50nm, (**b**) 75nm and (**c**) 100 nm. All measurements are at the minimum measured power for reset.

**Table S1 - Device dimensions from FIB-SEM micrograph. Layers arranged in the order they are displayed in Supplementary Fig. 1.**

| Nominal Via Size | Via diameter (nm) | Pt thickness (nm) | TiN thickness (nm) | GST thickness (nm) | $SiO_2$ thickness (nm) | Etched W depth (nm) | W thickness (nm) | Thermal $SiO_2$ thickness (nm) | Si Substrate (µm) |
|---|---|---|---|---|---|---|---|---|---|
| 50 nm | 25 | 65 | 10 | 50 | 10 | 15 | 40 | 100 | 4 (minimal thickness that does not change $\Delta T$ compared with real thickness of 400 µm) |
| 75 nm | 55 | 55 | 20 | 50 | 20 | 15 | 45 | 100 | |
| 100 nm | 90 | 95 | 25 | 50 | 25 | 25 | 60 | 100 | |
| 200 nm | 175 | 60 | 10 | 40 | 15 | 25 | 50 | 100 | |

**Table S2- Material properties** (*note GST and its surrounding are at high temperature during reset)

| Parameters \ Materials | **$SiO_2$** | **W** | **TiN** | **Pt** | **GST** | **Si** |
|---|---|---|---|---|---|---|
| Thermal conductivity (W/(m·K)) | 1.65 | 30 | 20 | 40 | 2 | 100 |
| Density(kg/($m^3$) | 2,200 | 19,350 | 21,450 | 21,450 | 6,200 | 2,329 |
| Heat capacity at constant pressure (J/(kg·K) | 730 | 132 | 133 | 133 | 220.7 | 700 |
| Electrical conductivity(S/m) | 1E-10 | 1E6 | 5E5 | 1E6 | 1E5 | 100 |
| TBR ($m^2$K/GW) | 10 | - | - | - | 18 | - |